\documentclass[conference,a4paper]{IEEEtran}

\usepackage{amssymb,amsmath,amsthm}
\usepackage{algorithm,algorithmic}
\usepackage{epsfig,graphics,graphicx,xcolor}
\usepackage{cite}
\usepackage{siunitx}
\usepackage[nolist]{acronym}
\usepackage[normalem]{ulem}
\usepackage[capitalize]{cleveref}
\usepackage{tikz}
\usetikzlibrary{shapes.geometric}
\usepackage{pgfplots}
\pgfplotsset{compat=1.16}

\newcommand{\transpose}{\mathsf{T}}

\newcommand{\setN}{\mathcal{N}}
\newcommand{\setS}{\mathcal{S}}

\newcommand{\abf}{\mathbf{a}}
\newcommand{\Abf}{\mathbf{A}}
\newcommand{\bbf}{\mathbf{b}}
\newcommand{\Bbf}{\mathbf{B}}

\newcommand{\gbf}{\mathbf{g}}
\newcommand{\Gbf}{\mathbf{G}}
\newcommand{\xbf}{\mathbf{x}}
\newcommand{\Xbf}{\mathbf{X}}

\newcommand{\onebf}{\boldsymbol{1}}

\newcommand{\Id}{\mathbf{I}}

\newcommand{\R}{\mathbb{R}}

\DeclareMathOperator{\supp}{supp}

\DeclareMathOperator*{\argmin}{arg\,min}

\newcommand{\rtt}{\ensuremath{\varepsilon}}
\newcommand{\offset}{\ensuremath{\rho}}
\newcommand{\gradcomm}{\ensuremath{\gamma}}

\newcommand{\smax}{\ensuremath{\overline{s}}}
\newcommand{\rvngc}{\ensuremath{T_{NGC}}}
\newcommand{\lindex}{\ensuremath{i}}

\newcommand{\define}{\ensuremath{\triangleq}}

\newcommand{\cardworkers}[1][\ntasks]{\ensuremath{Z}_{#1}^t}
\newcommand{\layer}{\ensuremath{u}}

\newcommand{\tp}{\ensuremath{t^\prime(\layer)}}
\newcommand{\tpp}{\ensuremath{t^{\prime \prime}(\layer)}}

\newcommand{\respworker}[1][\layer]{\ensuremath{T_i^{(#1)}}}
\newcommand{\failprob}{\ensuremath{p_{e}}}

\newtheorem{theorem}{Theorem}
\newtheorem{corollary}{Corollary}
\newtheorem{lemma}{Lemma}
\newtheorem{observation}{Observation}

\newtheorem{definition}{Definition}

\newcommand{\ie}{i.e., }
\newcommand{\eg}{e.g., }

\newcommand{\ExtVe}[1]{}
\newcommand{\btheta}{\ensuremath{\boldsymbol{\theta}}}
\newcommand{\lrate}{\ensuremath{\eta}}
\newcommand{\dm}{\ensuremath{\mathbf{D}}}

\newcommand{\hardstragglers}{\ensuremath{\kappa}}
\newcommand{\cdferl}[1][\sigma]{\ensuremath{F_{#1}(t)}}
\newcommand{\cdfTi}[1][\layer]{\ensuremath{F_{#1}(t)}}

\newcommand{\limit}[1][\layer]{\ensuremath{\hat{\lindex}_{#1}}}

\begin{acronym}
    \acro{gc}[GC]{gradient code}
    \acro{ngc}[NGC]{nested gradient code}
    \acro{iid}[i.i.d]{independent and identically distributed}
    \acro{cdf}[CDF]{cumulative density function}
\end{acronym}

\IEEEoverridecommandlockouts

\begin{document}

\title{{Nested Gradient Codes for Straggler Mitigation in Distributed Machine Learning}}
\author{
    \IEEEauthorblockN{Luis Maßny, Christoph Hofmeister, Maximilian Egger, Rawad Bitar and Antonia Wachter-Zeh}
    \IEEEauthorblockA{School of Computation, Information and Technology, Technical University of Munich (TUM), Munich, Germany\\\{\small{\texttt{luis.massny, christoph.hofmeister, maximilian.egger, rawad.bitar, antonia.wachter-zeh}}\}\texttt{@tum.de}}
    \thanks{
    This work was supported by the Bavarian Ministry of Economic Affairs, Regional Development and Energy within the scope of the 6G Future Lab Bavaria, by the Federal Ministry of Education and Research of Germany in the joint project 6G-life, project identification number, 16KISK002, and by the DFG (German Research Foundation) under Grant Agreement No. WA 3907/7-1.
    }
\vspace{-1cm}}

\maketitle

\begin{abstract}
We consider distributed learning in the presence of slow and unresponsive worker nodes, referred to as stragglers. In order to mitigate the effect of stragglers, gradient coding redundantly assigns partial computations to the worker such that the overall result can be recovered from only the non-straggling workers. Gradient codes are designed to tolerate a fixed number of stragglers. Since the number of stragglers in practice is random and unknown a priori, tolerating a fixed number of stragglers can yield a sub-optimal computation load and can result in higher latency. We propose a gradient coding scheme that can tolerate a flexible number of stragglers by carefully concatenating gradient codes for different straggler tolerance. By proper task scheduling and small additional signaling, our scheme adapts the computation load of the workers to the actual number of stragglers. We analyze the latency of our proposed scheme and show that it has a significantly lower latency than gradient codes.
\end{abstract}

\section{Introduction}\label{sec:Intro}
The training of state-of-the-art machine learning models requires high computational loads and operates on huge data sets, which may exceed the storage capacity of a single machine~\cite{lecunDeepLearning2015}. Distributing computations at a large scale is, therefore, inevitable in the model training. We consider the setting of synchronous distributed learning, in which a main node assigns partial computations to multiple worker nodes. It is well-known that large-scale distributed computation systems suffer from stragglers, \ie slow or unresponsive workers~\cite{deanTailScale2013}. Coding-theoretic solutions are proposed to mitigate the effect of stragglers through redundant computation assignment and treating the stragglers as erasures. The proposed encoding depends on the type of computation. For example, linear error-correcting codes are used to mitigate stragglers in matrix-vector multiplications, \eg\cite{leeSpeedingDistributedMachine2018,ramamoorthyUniversallyDecodableMatrices2019,severinsonBlockDiagonalLTCodes2019,maityRobustGradientDescent2019a}, and evaluation of polynomials, \eg \cite{yuLagrangeCodedComputing2019, yuPolynomialCodesOptimal2017}.

However, the training of common machine learning models such as deep neural networks may require the computation of highly non-linear gradients~\cite{lecunDeepLearning2015}. Gradient coding~\cite{tandonGradientCoding2017} is a coding strategy that allows straggler mitigation in distributed non-linear gradient computations. The main idea is to assign the partial computation to the workers using repetition or fractional repetition codes and apply linear error-correcting codes over the computed gradients at each worker. The main node is then guaranteed to obtain the sum of all computed gradients even if $s$ workers are stragglers. The required redundancy is shown to be at least $s+1$. Gradient coding is also related to the sum-recovery problem which arises in many learning algorithms, such as distributed, stochastic or batched gradient descent\footnote{Although our main motivation for sum-recovery are machine learning applications, sum-recovery codes can be applied to speed up the distributed computation of any non-linear function by considering their Taylor series expansion, as pointed out in~\cite{mallickRatelessCodesDistributed2021}.}. Several variations of the gradient coding framework are considered, \eg for partial recovery~\cite{sarmasarkarGradientCodingPartial2021}, for communication-efficiency~\cite{kadheCommunicationEfficientGradientCoding2020}, and with clustering~\cite{buyukatesGradientCodingDynamic2022}.

The previously mentioned works design codes for a particular number of stragglers. Straggling, however, is a stochastic process~\cite{deanTailScale2013}. In practice, the actual number of stragglers is unknown a priori. In particular, since gradient descent-based machine learning algorithms are iterative in nature, the number of stragglers may fluctuate between iterations. For this reason, flexible coding schemes have been developed, which adapt to the actual number of stragglers. These include flexible matrix-matrix multiplication~\cite{liFlexibleDistributedMatrix2022} and the use of rateless codes~\cite{mallickRatelessCodesDistributed2021}. While rateless codes can tolerate a flexible number of stragglers for sum-recovery, they require the replication of the whole data set to each worker, which is infeasible for many state-of-art machine learning tasks. Furthermore, they increase the communication overhead, since each worker transmits multiple encoded symbols, each of the same size as the model. For the case of persistent stragglers, the authors in~\cite{amiri2019computation} propose a careful scheduling of the compute tasks.

Approximate gradient coding is an approach specifically tailored to tolerate a flexible number of stragglers in distributed machine learning~\cite{wangErasureHeadDistributedGradient2019, bitarStochasticGradientCoding2020, sakorikarSoftBIBDProduct2022, charlesApproximateGradientCoding2017, glasgowApproximateGradientCoding2021,kosaianParityModelsErasurecoded2019}. The flexible straggler tolerance is guaranteed by allowing the main node to recover an approximate of the desired gradient computation if less workers than required for exact recovery respond.
Although approximate recovery decreases the computation latency per iteration, it penalizes the result by a reconstruction error, which slows down the convergence speed per iteration~\cite{bitarStochasticGradientCoding2020}. Approximate reconstruction rules, moreover, might perform poorly in some particular settings, such as non-identically distributed training data~\cite{chenRevisitingDistributedSynchronous2017, tandonGradientCoding2017}, and are incompatible with several advanced gradient descent techniques, such as momentum method~\cite{sutskeverImportanceInitializationMomentum}.

In this paper, we focus on distributed gradient descent with exact recovery. The goal is to reconstruct the overall gradient in the presence of a random number of stragglers with minimum computation overhead and small latency. We construct \emph{\acp{ngc}}, a code construction that simultaneously concatenates multiple \acp{gc} with different straggler tolerance. By jointly designing those \acp{gc} and carefully scheduling the computations at each worker, \acp{ngc} tolerate a flexible number of stragglers, guarantee a minimum computation load and reduce the latency.

We introduce the problem setting in more detail in \cref{sec:setting}. The code construction and the main result is then presented in \cref{sec:ngc}. In \cref{sec:waiting_time}, we analyze the latency of \acp{ngc} and conventional \acp{gc}. We also present simulation results that demonstrate that \acp{ngc} have superior performance over conventional \acp{gc}.

\section{Preliminaries and System Model}\label{sec:setting}
\subsection{Notation}
Matrices are denoted by bold upper case letters, \eg $\mathbf{Z}$ and the corresponding $i$-th row by bold lower case letters, \eg $\mathbf{z}_i$, respectively. Sets are written in calligraphic letters, \eg $\mathcal{Z}$. Let $n \in \mathbb{N}$, then we define $[n] \define \{1, 2, \cdots n\}$. We use $\supp\left( \mathbf{z} \right)$ to denote the support set of a vector $\mathbf{z}$ and $\left\| \mathbf{z} \right\|_0$ to denote its support size.

\subsection{Gradient descent}
Gradient descent is the main building block of several machine learning algorithms. It is used to iteratively find the minimum of a given loss function $L$ as defined next. Consider a data matrix $\dm\in \mathbb{R}^{m \times c}$ whose $m$ rows are called the training samples and $c$ columns are called the features. Let $\mathbf{y} \in \mathbb{R}^m$ be a column vector of labels, \ie $y_i$ is the label of $\mathbf{d}_i$, $i=1,\dots,m$. Define the function $f(\mathbf{d}_i,\boldsymbol{\theta})$ parametrized by a column vector $\boldsymbol{\theta} \in \mathbb{R}^\ell$ that aims to approximate $y_i$ for all $i=1,\dots,m$.
The goal is to find a $\boldsymbol{\theta}^\star$ that gives the best approximation $f(\mathbf{d}_i,\theta^\star)$ of $y_i$ for all $i \in [m]$.
To that end, the loss function $L(y_i,f(\mathbf{d}_i,\boldsymbol{\theta}))$ is defined to quantify how far is $f(\mathbf{d}_i,\boldsymbol{\theta})$ from $y_i$. We consider problems in which finding $\boldsymbol{\theta}^\star$ amounts to solving the following optimization problem%
\vspace{-1ex}
\begin{equation*}
    \btheta^\star = \argmin_{\btheta\in \R^\ell}\sum_{i=1}^{m} L\left(y_i, f(\mathbf{d}_i, \btheta)\right).
\end{equation*}

Gradient descent is an iterative algorithm used to solve this optimization problem. It first initializes the vector $\btheta_0 \in \mathbb{R}^\ell$ with random values. Then, at each iteration $t$, the vector $\btheta_t$ is updated as
\begin{equation}\label{eq:update_rule}
    \btheta_{t+1} = \btheta_{t} - \frac{\lrate_t}{m} \sum_{i=1}^m\nabla_{\btheta} L(y_i, f(\mathbf{d}_i, \btheta_t)) \triangleq \btheta_{t} - \frac{\lrate_t}{m} \mathbf{g}_{t},
\end{equation}
where the learning rate $\lrate_t$ is a parameter of the algorithm and $\nabla_{\btheta} L({y}_i, f(\mathbf{d}_i, \btheta_t))$ is the gradient of the loss function evaluated at $y_i$, $\mathbf{d}_i$, and $\btheta_t$. The algorithm runs until a satisfactory model $\btheta$ is obtained.

\subsection{Coded distributed dradient descent}
We consider a distributed computing cluster consisting of a main node and $n$ workers. The goal is to distributively compute the vector $\mathbf{g}_t$, termed \emph{full gradient}, at each iteration $t$.
The data matrix is partitioned into $k$ disjoint row block matrices\footnote{We assume that $k$ divides $m$. Otherwise, the data can be zero-padded.} $\dm_i, i=1,\dots,k,$ each with dimension $\frac{m}{k} \times \ell$. Therefore, the vector $\mathbf{g}_t$ is composed of $k$ \emph{partial gradients} defined as $\mathbf{g}_{i,t} \triangleq \displaystyle \sum_{\mathbf{d}_j \in \dm_i}\nabla_{\btheta} L({y}_i, f(\mathbf{d}_j, \btheta_t))$. 

We tackle the problem of exactly computing the full gradient $\mathbf{g}_t$ in the presence of a random number of stragglers $s \in [n]$. It is shown in \cite[Theorem 1]{tandonGradientCoding2017} that in this setting each partial gradient has to be computed by at least $s+1$ workers. In particular, if all workers have the same computation load, then each worker computes at least $\frac{k}{n}\left( s+1 \right)$ partial gradients. In practical scenarios, the number $\smax$ of maximum stragglers that can be tolerated is limited by a storage constraint. Let $\beta$ be the number of rows from $\dm$ that a worker can store. Then the number of row blocks of size $\frac{c}{m}$ that a worker can store is limited to $\lfloor \frac{\beta m}{c} \rfloor$. Therefore, \mbox{$\smax \leq \lfloor \frac{n \beta m }{k c} \rfloor - 1$} holds. The computation of $\frac{k}{n}$ gradients is defined as an atomic task. As we assume $k=n$ for the sake of simplicity, an atomic task is a partial gradient computation. The results can be generalized by assigning $\frac{k}{n}$ partial gradient computations jointly.

At the beginning of the algorithm, the main node assigns the row block matrices redundantly to the workers. At each iteration, the main node sends $\btheta_t$ to the workers. The workers compute the partial gradients on their assigned data and send a linear combination of the computed partial gradients to the main node. The data assignment and the linear combination of the computed gradients at each iteration should ensure that the main node can obtain the full gradient in the presence of $s\in [\smax]$ stragglers. In the sequel, we focus at only one iteration of the algorithm, and omit the dependence on $t$, as the coding ideas remain the same across iterations. The response of each non-straggling worker $i$ is denoted by $\xbf_i \in \R^\ell$, which is a function of the assigned partial gradients. The responses of the workers are defined by a generator matrix $\Bbf$ as
\begin{equation*}
\Xbf = \Bbf \Gbf,
\end{equation*}
where $\Gbf = \left( \gbf_1,\dots,\gbf_k \right)^\transpose \in \R^{k \times \ell}$.

\subsection{Model of the workers response time}
Our main figure of merit is the \emph{latency} $T$ of the system, \ie the time required for one iteration of gradient descent.
We model the service time for a single task on worker $i$ as an \ac{iid} random variable, for which we consider the shifted exponential distribution model, \eg \cite{liang2014tofec,leeSpeedingDistributedMachine2018}, with shift $\offset$ and rate $\lambda$. The shift captures the deterministic part of the computation time, and the tail of the exponential describes stochastic delays \cite{severinsonBlockDiagonalLTCodes2019}.
We assume a homogeneous compute cluster in which $\offset$ and $\lambda$ are common for all workers. The transmission of the model and the gradients is modeled by a constant \emph{communication delay} $\gradcomm$. In the sequel, we will analyze schemes that require an additional lightweight signaling overhead in the order of few bits, which is captured by the constant $\rtt$. In addition to stochastic delays, stragglers may fail completely with probability $\failprob$ independently from each other. For a non-failing worker, the time $T_i^{(\layer)}$ required to compute $\layer$ tasks is given by
\begin{equation}
    \label{eq:resptime}
    \respworker =
    \gradcomm + \rtt + \layer \offset + \tau_i^{(\layer)},
\end{equation}
where $\tau_i^{(\layer)}$ is the sum of $\layer$ \ac{iid} exponential random variables, thus following an Erlang distribution with shape $\layer$ and rate $\lambda$. For the random variable $\respworker$, we obtain the following \ac{cdf} for $t \geq \gradcomm + \rtt + \layer \offset$ as
\newcommand{\auxvar}{\ensuremath{\tau}}
\begin{equation}
    \label{eq:layer_cdf}
    \cdfTi[\layer] = 1 \! - \!\! \sum_{k = 0}^{\layer} \frac{\lambda^k}{k!} \left( t - (\gradcomm + \rtt + \layer \offset) \right)^k e^{-\lambda (t- (\gradcomm + \rtt + \layer \offset))}.
\end{equation}
For $t < \gradcomm + \rtt + \layer \offset$, we have $\cdfTi[\layer] = 0$.
We assume that the main node is aware of the progress at every worker, \ie the number of tasks completed per worker, at any times. Furthermore, the main node can send a termination message containing a small integer value. The additional communication is reasonable, since it can be realized by some lightweight signaling in the order of a few bits, which is negligible compared to the size of the gradients. The small overhead of the termination message may also be acceptable, since it can preserve the anonymity of the individual workers' progress by its broadcast nature.

\section{Nested Gradient Codes}\label{sec:ngc}
In this section, we present our construction of \ac{ngc} that is robust against a \emph{flexible} number of stragglers. This construction uses multiple \acp{gc} that are robust against a \emph{fixed} number of stragglers $\sigma$ as introduced in \cite{tandonGradientCoding2017} and defined as follows.

\begin{definition}[Gradient code \cite{tandonGradientCoding2017}]
\label{def:regular_gc}
An $\left( n,k,\sigma \right)$ gradient code is defined by a tuple of matrices $\left( \Abf_\sigma, \Bbf \right)$, such that
\begin{itemize}
    \item $\Abf_\sigma \in \R^{ \binom{n}{\sigma} \times n }$, $\Bbf_\sigma \in \R^{n \times k}$
    \item every $n-\sigma$ rows of $\Bbf_\sigma$ span an $(n-\sigma)$-dimensional subspace $\setS$, where $\onebf_{1 \times k} \in \setS$,
    \item $\left\| \bbf_j \right\|_0 \geq \frac{k}{n}\left( \sigma+1 \right)$ for every row $j=1,\dots,n$,
    \item for any subset $\setN_j \subseteq \left\{ 1,\dots,n \right\}$ of size $n-\sigma$, there exists a row $\abf_j$ that has $\supp\left( \abf_j \right) \subseteq \setN_j$,
    \item $\Abf \Bbf = \onebf_{\binom{n}{\sigma} \times k}$.
\end{itemize}
\end{definition}

A \ac{gc} as in \cref{def:regular_gc} tolerates up to $\sigma$ stragglers~\cite[Lemma 1]{tandonGradientCoding2017}.
If the actual number of stragglers $s$ is smaller than the design parameter $\sigma$, the computations of $\sigma - s$ workers will be ignored. Equivalently, every non-straggling worker has a computation overhead of at least a factor $\frac{\sigma+1}{s+1}$ compared to a \ac{gc} designed for $\sigma=s$. To avoid this overhead, we propose the concatenation and careful nesting of \acp{gc} with different values of $\sigma$. Our main observation is that an $(n,k,\sigma+1)$ \ac{gc} can operate on the same partial gradients as an $(n,k,\sigma)$ \ac{gc}. Formally, the row support of the encoding matrix $\Bbf_{\sigma+1}$ for the first \ac{gc} is a superset of the row support of the encoding matrix $\Bbf_{\sigma}$ for the second \ac{gc}.

We call a set of $\smax+1$ different $(n, k, \sigma)$ \acp{gc} for $\sigma=0, \dots, \smax$ that satisfies this property an $(n,k,\smax)$ \ac{ngc}. In Definition~\ref{def:nested_gc}, we formally define an \ac{ngc} construction that can tolerate any number of $s \leq \smax$ stragglers. 

\begin{definition}[Nested gradient code]
\label{def:nested_gc}
An $\left( n,k,\smax \right)$ nested gradient code is defined by a set of tuples $\left\{ \left( \Abf_\sigma, \Bbf_\sigma \right), \, \sigma=0,\dots,\smax \right\}$, referred to as component gradient codes, such that for every $\sigma=0,\dots,\smax$, it holds that
\begin{itemize}
    \item $\left( \Abf_\sigma, \Bbf_\sigma \right)$ is an $( n,k,\sigma )$ gradient code according to \cref{def:regular_gc},
    \item every row $\bbf_\sigma^{(i)}$ has $\supp\left( \bbf_{\sigma}^{(i)} \right) \subseteq \supp\left( \bbf_{\sigma+1}^{(i)} \right)$ for \mbox{$\sigma < \smax$}.
\end{itemize}
\end{definition}

\begin{observation}[Straggler tolerance of nested gradient codes]
An $\left( n,k, \smax \right)$ \ac{ngc} allows the main node to tolerate any number of $s=0,1,\dots,\smax$ stragglers with each non-straggler having at most the computation load of a \ac{gc} with $\sigma = s$. The per-worker storage requirement of an $\left( n,k, \smax \right)$ \ac{ngc} is the same as that of an $\left( n,k, \sigma = \smax \right)$ \ac{gc}, but extra signaling is needed between the workers and the main node to keep track of the progress of the workers. 
\end{observation}

In order to leverage the nested support structure of the encoding matrices, we define a particular task scheduling at each worker. We let each worker start with the partial gradient computation for the component \ac{gc} for $\sigma=0$, and let it proceed with the partial gradient computation for larger values of $\sigma$ successively. Due to the layered task scheduling, we refer to the partial gradient computations for the component \ac{gc} for $\sigma$ as the \emph{$\sigma$-th computation layer}.
\vspace{-0.5em}
\begin{figure}[H]
    \centering
    \resizebox{\linewidth}{!}{
\begin{tikzpicture}

\definecolor{darkcyan0107164}{RGB}{0,107,164}
\definecolor{darkgray171}{RGB}{171,171,171}
\definecolor{darkgray176}{RGB}{176,176,176}
\definecolor{darkorange25512814}{RGB}{255,128,14}
\definecolor{lightgray204}{RGB}{204,204,204}

\begin{axis}[
height=8cm,
width=1.5\linewidth,
legend cell align={center},
legend columns=3,
legend style={
  at={(0.5,-0.2)},
  /tikz/every even column/.append style={column sep=0.5cm},
  anchor=north,
  draw=none 
},
tick align=outside,
tick pos=left,
x grid style={darkgray176},
xlabel={\(\displaystyle t\)},
xmajorgrids,
xmin=2, xmax=18,
xtick style={color=black},
y grid style={darkgray176},
ylabel={\(\displaystyle P(T \leq t)\)},
ymajorgrids,
ymin=0, ymax=1,
ytick style={color=black}
]
\addplot [ultra thick, darkcyan0107164, dotted]
table {sigma_0.csv};
\addlegendentry{No coding, $\sigma = 0$}
\addplot [ultra thick, darkorange25512814, dashed]
table {sigma_3.csv};
\addlegendentry{GC with $\sigma = 3$}
\addplot [ultra thick, darkgray171]
table {smax_3.csv};
\addlegendentry{NGC with $\smax=3$}
\end{axis}

\end{tikzpicture}
    }
    \caption{
    Latency of a \ac{ngc} with $\smax=3$ and a \ac{gc} with $\sigma=3$, both for $n=k=8$.
    }
    \label{fig:comp}
\end{figure}
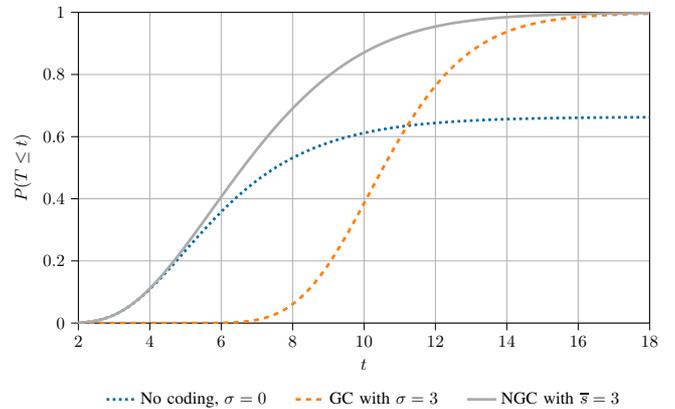
\vspace{-0.5em}
The advantage of \acp{ngc} over \acp{gc} according to \cref{def:regular_gc} is illustrated in \cref{fig:comp}, which shows the reduced latency of \acp{ngc}. The probability that all workers respond is $(1-\failprob)^n$. Thus, for $n=k=8$ workers and a failure probability of $\failprob=0.05$, the curve for $\sigma=0$ tends to $\lim_{t\rightarrow \infty} P(T\leq t) \approx0.66$. Considering a \ac{gc} and an \ac{ngc} with parameters $\sigma=\smax=3$, \ie each row block matrix is replicated at $4$ workers, the \ac{ngc} clearly outperforms the \ac{gc} given that the signaling time $\rtt$ is negligible.
An analysis of the latency of \ac{ngc} and a thorough comparison of \acp{gc} and \acp{ngc} are carried out in \cref{sec:waiting_time}.

We provide an explicit construction for \acp{ngc} in Algorithm~\ref{alg:construction}. 
It uses the algorithm $\texttt{determineCoef()}$ from \cite[Algorithm 2]{tandonGradientCoding2017} as a subroutine to construct encoding matrices $\Bbf_\sigma$ with nested cyclic repetition support structure and the algorithm $\texttt{findDecoding()}$ from \cite[Algorithm 1]{tandonGradientCoding2017} to construct the corresponding decoding matrices $\Abf_\sigma$. Due to space limitations, we refer to the reference for further explanations. As shown in Lemma~\ref{lma:construction}, this construction yields a valid \ac{ngc}.

\begin{algorithm}
\caption{Construction of an $\left( n,k,\smax \right)$ nested gradient code.}
\label{alg:construction}
\begin{algorithmic}
\REQUIRE{ $n,k,\smax$ }
\REQUIRE{ \tt{findDecoding()}, \tt{determineCoef()} }

\STATE{ $\Bbf_0 \gets \Id_{n \times k}$ }
\STATE{ $\Abf_0 \gets \onebf_{1 \times n}$ }
\FOR{ $\sigma = 1,\dots,\smax$ }
    \STATE{ $\Bbf_\sigma \gets$ \tt{determineCoef}($n$,$\sigma$) }
    \STATE{ $\Abf_\sigma \gets$ \tt{findDecoding}($\Bbf_\sigma$,$\sigma$) }
\ENDFOR
\RETURN{ $\left\{ \left( \Abf_\sigma, \Bbf_\sigma \right), \, \sigma=0,\dots,\smax \right\}$ }
\end{algorithmic}
\end{algorithm}

\begin{lemma}[Cyclic repetition construction]
\label{lma:construction}
Algorithm~\ref{alg:construction} generates an $\left(n, k, \smax \right)$ nested gradient code.
\begin{proof}
By~\cite[Theorem 3]{tandonGradientCoding2017}, Algorithm~\ref{alg:construction} creates an $\left( n,k,\sigma \right)$ \ac{gc} with encoding matrix $\Bbf_\sigma$ and decoding matrix $\Abf_\sigma$ for $\sigma=1,\dots,\smax$. Furthermore, since $\Bbf_0 = \Id_{n \times k}$ and $\Abf_0 \Bbf_0 = \onebf_{1 \times n} \Id_{n \times k} = \onebf_{1 \times k}$, according to Definition~\ref{def:regular_gc}, the tuple $\left(\Bbf_0,\Abf_0\right)$ is a valid encoding matrix for an $\left( n,k,0 \right)$ \ac{gc}. Finally, due to the cyclic repetition structure of matrices generated by \texttt{determineCoef()}, we have that $\supp\big( \bbf_\sigma^{(i)} \big)=\left\{ i,i+1,\dots,i+\sigma+1 \right\}$ and $\supp\big( \bbf_{\sigma+1}^{(i)} \big)=\left\{ i,i+1,\dots,i+(\sigma+1)+1\right\}$, such that $\supp\big(\bbf_{\sigma}^{(i)}\big) \subset \supp\big(\bbf_{\sigma+1}^{(i)}\big)$.
\end{proof}
\end{lemma}

We note that in order to select one of the $\smax+1$ component \acp{gc} adaptively, we have to allow lightweight signaling between the workers and the main node. As argued before, the communication cost of this signaling is negligible compared to the the communication cost of the final responses. In particular, each worker can signal the completion of a gradient computation by a binary flag. The delay caused by the round-trip time of the signal does not cause a computation delay, since the workers can meanwhile continue with subsequent gradient computations in an asynchronous manner. Solely the final signaling from the workers to the main node and the termination signal from the main node to the workers causes a total delay of a single round trip time, which is orders of magnitudes smaller than the compute times and the time required for transmitting the final responses.

\section{Latency of \acp{ngc}}\label{sec:waiting_time}
We compare the latency $T_{GC}$ of \acp{gc} and $T_{NGC}$ of \acp{ngc}. The latency of the former under the considered model of workers response time is given in \cref{prop:cdf_rgc}. One of our main results is the latency analysis for \acp{ngc}, which is stated in Theorem~\ref{thm:wt_ngc}.

\begin{lemma}[Latency of a gradient code]
\label{prop:cdf_rgc}
The probability that the latency $T_{GC}$ of an $\left( n,k,\sigma \right)$ gradient code is less than or equal to $t$ is given by
\begin{align*}
    P(T_{GC} \leq t) &= \sum_{\hardstragglers=0}^{n} P(T_{GC} \leq t \mid \hardstragglers) P(\hardstragglers) \\
    &= \sum_{\hardstragglers=0}^{\sigma} P(T_{GC} \leq t \mid \hardstragglers) \binom{n}{\hardstragglers} \failprob^\hardstragglers (1-\failprob)^{n-\hardstragglers},
\end{align*}
where $\hardstragglers$ is the number of completely unresponsive stragglers. The conditional \ac{cdf} for $\hardstragglers \leq \sigma$ and $n^\prime = n-\hardstragglers$ non-failing workers is given by
\begin{equation*}
    P(T_{GC} \leq t \mid \hardstragglers) = \!\!\! \sum_{\tau=n-\sigma}^{n^\prime} \!\! \binom{n^\prime}{\tau} \! \left[ \cdfTi[\sigma+1] \right]^{\tau} \! \left[ 1-\cdfTi[\sigma+1] \right]^{n^\prime-\tau}.
\end{equation*}
\begin{proof}
The proof follows directly by marginalizing over the number of tolerable failing workers $\hardstragglers \leq \sigma$, and applying the \ac{cdf} for the $(n-\sigma)$-th order statistic.
\end{proof}
\end{lemma}
\begin{theorem}[Latency of a nested gradient code]
\label{thm:wt_ngc}
The probability that the latency $T_{NGC}$ of an $\left( n,k,\smax \right)$ nested gradient code is less than or equal to $t$ is given by
\begin{align*}
    P(T_{NGC} \leq t) &= \sum_{\hardstragglers=0}^{\smax} P(T_{NGC} \leq t \mid \hardstragglers) P(\hardstragglers) \\
    &= \sum_{\hardstragglers=0}^{\smax} P(T_{NGC} \leq t \mid \hardstragglers) \binom{n}{\hardstragglers} \failprob^\hardstragglers (1-\failprob)^{n-\hardstragglers},
\end{align*}
where the conditional probability for $\hardstragglers \leq \smax$ failing workers is
\begin{align*}
&P(\rvngc \leq t \mid \hardstragglers) = \\
&1 -
\sum_{\lindex_{\smax+1} = 0}^{\limit[\smax+1]} \!\!\!\! \cdots \!\! \sum_{\lindex_{1} = 0}^{\limit[1]}
\binom{n-\hardstragglers}{\lindex_{1},\dots,\lindex_{\smax+1}}
\left(\cdfTi[\smax+1]\right)^{\lindex_{\smax+1}} \left(1 - \cdfTi[1]\right)^{\limit[0]} \nonumber \\
& \prod_{\layer = 1}^{\smax} \big(\cdfTi[\layer] - \cdfTi[\layer+1]\big)^{\lindex_\layer}.
\end{align*}
and we define
$\limit \define
\min\bigg\{
    n-\layer, \, n-\hardstragglers-{ \sum_{v=\layer+1}^{\smax+1} \lindex_v}
\bigg\}$.

\begin{proof}
We first marginalize over the number of failing workers $\hardstragglers$ that can be handled by the \ac{ngc}, \ie $\hardstragglers\leq\smax$. Given that $\hardstragglers$ workers fail completely, we express the probability that $\rvngc \leq t$ as the counter probability of the event that we \emph{cannot} decode in any of the component $(n,k,\sigma)$ \acp{gc}, $\sigma \in \{0, \dots, \smax\}$, up to time $t$.

Recall that in order to be able to decode from layer $\layer \in \{ 0,\dots,\smax+1 \}$, we require $n-\layer+1$ workers to finish layer $\layer$. Let $\lindex_\layer$ denote the number of that have finished exactly $\layer$ layers yet. Then we are not able to decode if $\lindex_\layer \leq n-\layer-\sum_{v=\layer+1}^{\smax+1} \lindex_v$. The idea is now to sum over the probabilities that we are not able to decode for all possible realizations of $\lindex_0,\dots,\lindex_{\smax+1}$. Since we have $\hardstragglers$ failing servers, it is furthermore, not possible to have layer $\layer$ finished at more than $n-\hardstragglers-\sum_{v=\layer+1}^{\smax+1} \lindex_v$ workers. In total, we recursively define the limit for $\lindex_\layer$ as
$\limit \define
\min\bigg\{
    n-\layer, \, n-\hardstragglers-{ \sum_{v=\layer+1}^{\smax+1} \lindex_v}
\bigg\}$. A particular realization $\lindex_0,\dots,\lindex_{\smax+1}$ occurs with probability
\begin{equation}
\label{eq:prob_realization}
\prod_{\layer=0}^{\smax+1} P\left( \cardworkers[\layer] = \lindex_\layer 
\middle| \sum_{\tau=\layer+1}^{\smax+1} \cardworkers[\tau] = \sum_{\tau=\layer+1}^{\smax+1} \lindex_{\tau} \right)
\end{equation}
where $\cardworkers[\layer]$ is a random variable, denoting the number of non-failing workers who have finished exactly $\layer$ layers within $t$.

It remains to determine the above probabilities for all layers $\layer \in \{0, \dots, \smax+1\}$. For layer $\layer = \smax+1$, the probability that $\cardworkers[\smax+1] = \lindex_{\smax+1}$ is given by
\vspace{-0.5em}
\begin{equation*}
    P\left( \cardworkers[\smax+1] = \lindex_{\smax+1} \right) = \binom{n-\hardstragglers}{\lindex_{\smax+1}} \left(\cdfTi[\smax+1]\right)^{\lindex_{\smax+1}}
\end{equation*}
since all workers have \ac{iid} response times according to \eqref{eq:resptime}.
\renewcommand{\tp}{\ensuremath{t^\prime}}
\renewcommand{\tpp}{\ensuremath{t^{\prime\prime}}}
For layer $\layer \in [\smax]$ and given that $\lindex_{\layer+1}$ workers finished exactly $\layer+1$ tasks, we have
\begin{align}
    &P\left(\cardworkers[\layer] = \lindex_\layer \middle| \sum_{\tau=\layer+1}^{\smax+1} \cardworkers[\tau] = \sum_{\tau=\layer+1}^{\smax+1} \lindex_{\tau} \right) \label{eq_line:middle_layers} \\
    &= \binom{n-\hardstragglers-\lindex_{\smax+1}-\dots-\lindex_{\layer+1}}{\lindex_\layer} P\big(\respworker[\layer] \leq t < \respworker[\layer+1]\big)^{\lindex_\layer} \nonumber \\
    &= \binom{n-\hardstragglers-\lindex_{\smax+1}-\dots-\lindex_{\layer+1}}{\lindex_\layer} \big(\cdfTi[\layer] - \cdfTi[\layer+1]\big)^{\lindex_\layer} \nonumber.
\end{align}
For $\layer=0$, we finally obtain
\begin{align*}
    P\left( \cardworkers[0] = \lindex_0 \middle| \sum_{\tau=1}^{\smax+1} \cardworkers[\tau] = \sum_{\tau=1}^{\smax+1} \lindex_{\tau} \right) &= P(\respworker[1] > t)^{\limit[0]} \\
    &= \left(1 - \cdfTi[1]\right)^{\limit[0]}.
\end{align*}

Then, we can conclude the proof by plugging the probabilities into \eqref{eq:prob_realization} and summing over all possible realizations for $\lindex_0,\dots,\lindex_{\smax+1}$.
\end{proof}
\end{theorem}

In case the constant computation time is $\offset=0$, \ie the time for computing a tasks is purely stochastic, \cref{thm:wt_ngc} simplifies to \cref{cor:wt_ngc}.
\begin{corollary} \label{cor:wt_ngc}
For $\offset=0$, we have from \cref{thm:wt_ngc} that
\vspace{-0.5em}
\begin{align*}
P(\rvngc \leq t \mid \hardstragglers) =& 1 -
\sum_{\lindex_{\smax+1} = 0}^{\limit[\smax+1]} \!\!\!\! \cdots \!\! \sum_{\lindex_{1} = 0}^{\limit[1]}
\binom{n-\hardstragglers}{\lindex_{1},\dots,\lindex_{\smax+1}}\\
& \left(e^{\lambda t} \right)^{\lindex_0} \prod_{\layer = 1}^{\smax} \left( \frac{1}{\layer!} e^{-\lambda t} (\lambda t)^\layer \right)^{\lindex_\layer} \!\!\! \cdferl[\smax]^{\lindex_{\smax+1}}.
\end{align*}
\begin{proof}
We can write
\begin{align*}
\cdfTi[\layer] - \cdfTi[\layer+1] &= \sum_{\nu = 0}^{\layer} \frac{1}{\nu !} e^{-\lambda\tpp} (\lambda t)^\nu -\sum_{\nu = 0}^{\layer-1} \frac{1}{\nu !} e^{-\lambda \tp} (\lambda t)^\nu\\
&= \frac{1}{\layer!} e^{-\lambda t} (\lambda t)^\layer
\end{align*}
and thus, \eqref{eq_line:middle_layers} in the proof of \cref{thm:wt_ngc} simplifies to
\vspace{-0.5em}
\begin{align*}
    & P\bigg(\cardworkers[\layer] = \lindex_\layer \bigg\vert \sum_{\tau=\layer+1}^{\smax+1} \cardworkers[\tau] = \sum_{\tau=\layer+1}^{\smax+1} \lindex_{\tau} \bigg) \\
    =& \binom{n-\hardstragglers-\lindex_{\smax+1}-\dots-\lindex_{\layer+1}}{\lindex_\layer} \bigg(\! \frac{1}{\layer!} e^{-\lambda t} (\lambda t)^\layer \bigg)^{\lindex_\layer},
\end{align*}
which concludes the proof.
\end{proof}
\end{corollary}

We now compare the latency of the proposed \ac{ngc} construction for $n=8$ and different design parameters $\smax$ to \acp{gc} from \cite{tandonGradientCoding2017} with different values of $\sigma$. We consider a rate of $\lambda=0.5$, a constant computation time of $\offset=0.5$ and a round-trip time of $\rtt=0.1$. Since the communication time $\gradcomm$ constitutes a fixed overhead for all schemes, we plot $t-\gradcomm$. We further set the failure probability per worker to $\failprob=0.05$. The resulting latency for these parameters are depicted in \cref{fig:comp}. 
In principle, one can see that the larger the straggler tolerance $\sigma$ or $\smax$ of the scheme, the larger $\lim_{t\rightarrow \infty} P(T\leq t)$. However, for \acp{gc}, choosing larger $\sigma$ causes delays. This does not hold for \acp{ngc}, where larger values for $\smax$ always come with smaller delay. As expected, the latency for the \acp{ngc} with $\smax$ is less than for a \ac{gc} with $\sigma=\smax$, given that the signaling in the order of one round-trip time is negligible. Beyond that, the \acp{ngc} for $\smax=4$ and $\smax=6$ almost have equal characteristics, which implies that the cost of storage at the workers can significantly be reduced at almost no cost in processing speed.
\newcommand{\tmp}{\ensuremath{\kappa}}

\vspace{-0.5em}
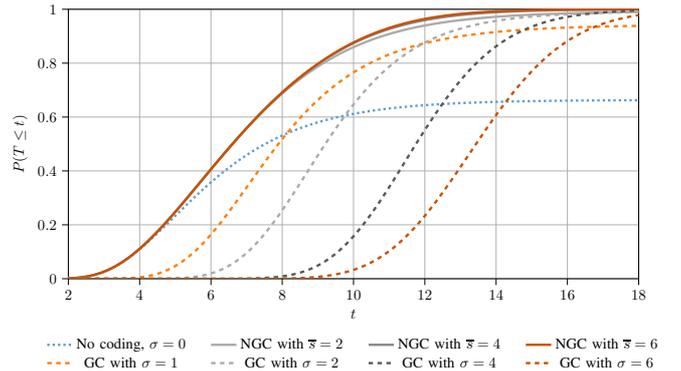
\begin{figure}[H]
    \centering
    \resizebox{\linewidth}{!}{
\begin{tikzpicture}

\definecolor{chocolate200820}{RGB}{200,82,0}
\definecolor{cornflowerblue95158209}{RGB}{95,158,209}
\definecolor{darkcyan0107164}{RGB}{0,107,164}
\definecolor{darkgray171}{RGB}{171,171,171}
\definecolor{darkgray176}{RGB}{176,176,176}
\definecolor{darkorange25512814}{RGB}{255,128,14}
\definecolor{dimgray89}{RGB}{89,89,89}
\definecolor{gray137}{RGB}{137,137,137}
\definecolor{lightgray204}{RGB}{204,204,204}
\definecolor{lightsteelblue162200236}{RGB}{162,200,236}

\begin{axis}[
height=8cm,
width=1.7\linewidth,
legend cell align={center},
legend columns=4,
legend style={
  at={(0.5,-0.2)},
  /tikz/every even column/.append style={column sep=0.5cm},
  anchor=north,
  draw=none 
},
tick align=outside,
tick pos=left,
x grid style={darkgray176},
xlabel={\(\displaystyle t\)},
xmajorgrids,
xmin=2, xmax=18,
xtick style={color=black},
y grid style={darkgray176},
ylabel={\(\displaystyle P(T \leq t)\)},
ymajorgrids,
ymin=-0, ymax=1,
ytick style={color=black}
]
\addplot [ultra thick, cornflowerblue95158209, dotted]
table {sigma_0.csv};
\addlegendentry{No coding, $\sigma = 0$}
\addplot [ultra thick, darkgray171]
table {smax_2.csv};
\addlegendentry{NGC with $\smax=2$}
\addplot [ultra thick, gray137]
table {smax_4.csv};
\addlegendentry{NGC with $\smax=4$}
\addplot [ultra thick, chocolate200820]
table {smax_6.csv};
\addlegendentry{NGC with $\smax=6$}

\addplot [ultra thick, darkorange25512814, dashed]
table {sigma_1.csv};
\addlegendentry{GC with $\sigma = 1$}
\addplot [ultra thick, darkgray171, dashed]
table {sigma_2.csv};
\addlegendentry{GC with $\sigma = 2$}
\addplot [ultra thick, dimgray89, dashed]
table {sigma_4.csv};
\addlegendentry{GC with $\sigma = 4$}
\addplot [ultra thick, chocolate200820, dashed]
table {sigma_6.csv};
\addlegendentry{GC with $\sigma = 6$}
\end{axis}

\end{tikzpicture}
    }
    \caption{
    \Acp{cdf} for the computation latency $T_{NGC}$ of \acp{ngc} and $T_{GC}$ of conventional \acp{gc} with $n=k=8$, different straggler tolerance $\sigma$ for \acp{gc} and maximum possible straggler tolerance $\smax$ for \acp{ngc}.
    }
    \label{fig:comp?full}
\end{figure}

\vspace{-0.5em}
\section{Conclusion and Future Directions}\label{sec:conc}
We proposed a nested gradient code construction to tackle the problem of stragglers in distributed gradient descent when the actual number of stragglers is not known a priori. We give a construction that adapts the computation load to the actual number $s \leq \smax$ of stragglers. By limiting the number of maximum tolerable stragglers $\smax$, the scheme can be tailored to a specified storage constraint. We analyze the latency of the proposed scheme and compare to a known gradient code construction with fixed straggler tolerance $\sigma$. Future research directions include the elimination of the signaling and the analytical investigation of the trade-off between the expected latency and the storage overhead in relation to the system parameters $\lambda$, $\offset$, $\gradcomm$, and $\rtt$.

\bibliographystyle{IEEEtran}

\end{document}